\documentclass[twoside]{article}
\usepackage{wrapfig}
\usepackage{graphicx}
\usepackage[T2A]{fontenc}
\usepackage[cp866nav]{inputenc}
\usepackage{amssymb}
\usepackage{cmpj2e}

\title[ Gauge Dependence of the Critical Dynamics at the Superconducting Transition]%
{Gauge Dependence of the Critical Dynamics at the Superconducting Transition%
}
\author[M. Dudka\refaddr, R. Folk\refaddr, G. Moser]
       {M. Dudka\refaddr{label1,label2}, R. Folk\refaddr{label2}, G. Moser\refaddr{label3}}
\addresses{
\addr{label1} Institute for Condensed Matter Physics, National
Academy of Sciences of Ukraine, UA--79011 Lviv, Ukraine.
\addr{label2} Institut f\"ur Theoretische Physik, Johannes Kepler
Universit\"at Linz,  A--4040 Linz, Austria
 \addr{label3} Institut f\"ur Physik und Biophysik, Universit\"at
Salzburg, A--5020 Salzburg, Austria}
\begin{document}

\maketitle

\begin{abstract}
The critical dynamics of superconductors in the charged regime is
reconsidered within field-theory. For the dynamics the
Ginzburg-Landau model with complex order parameter coupled to the
gauge field suggested earlier [Lannert et al. Phys. Rev. Lett.
{\bf 92}, 097004 (2004)] is used. Assuming relaxational dynamics
for both quantities the renormalization group functions within one
loop approximation are recalculated for different choices of the
gauge. A gauge independent result for the divergence of the
measurable electric conductivity is obtained only at the weak
scaling fixed point unstable in one loop order where the time
scales of the order parameter and the gauge field are different.
\keywords critical dynamics, superconducting transition
\pacs 05.70.Jk, 64.60.Ak, 64.60.Ht, 74.20.-z
\end{abstract}

\section{Introduction}
The nature of the static phase transition in superconductors was
an open question for decades, since due to the large correlation
length of existing superconducting materials the effect of
critical fluctuations was hard to observe in the vicinity of the
critical temperature $T_c$. Moreover it was unclear if the phase
transition is of first or second order and, if it is of second
order, to which universality class the transition belongs.  The
appearance of high-$T_c$ superconducting compounds with short
correlation lengths \cite{Lobb87} made the critical region of
superconductors experimentally accessible. In turn this leads to
comparison of the experiment with analytical results in order to
establish the critical properties of superconductors and their
universality class.

From the theoretical point of view the static critical properties
are now well understood. The theoretical model for the description
of the static critical properties was formulated in
\cite{Halperin74} and contains,  besides the two component ($n=2$)
order parameter (OP) and its fourth order interaction term, a
minimal coupling to a gauge field (GF) due to the charged
character of the OP (macroscopic wave function of the Cooper
pair). The coupling of the OP to the GF introduces an essential
difference to the superfluid phase transition, where the OP is
uncharged.

While for type-I superconductors fluctuation effects  are weak and
a mean-field analysis can be applied, the situation with type-II
superconductors is more complicated, since here fluctuations of
the OP cannot be neglected. The first renormalization group (RG)
analysis in  one-loop approximation \cite{Halperin74} lead to the
conclusion that a stable fixed point (FP) and therefore a second
order phase transition  only exists for OP dimensions $n$ larger
than 365.9. For OP dimension $n=2$ no stable FP was found and the
runaway solution was interpreted as a weak first order phase
transition. A two loop order  calculation of the renormalization
group equations within field theory indicated the possibility of a
continuous phase transition for $n=2$ \cite{Kolnberger90} if
certain resummations are performed. This has been investigated
further and led to the following picture \cite{Folk98}: There are
four FPs, two uncharged ones known from the standard
Ginzburg-Landau-Wilson (GLW) model and two charged ones. Depending
on the initial (background) conditions for the flow of the fourth
order coupling and the coupling to the charge (defining the value
of the Ginzburg parameter $\kappa$) one obtains a runaway flow or
the flow reaches the stable  charged FP. The other charged FP is
reached starting on the separatrix (defined by $\kappa=\kappa_c$)
separating the attraction region of the charged FP from the
runaway region. Physically this FP reached on the separatrix
describes tricritical behavior indicating that a tricritical point
separates the first order transition of the superconductors of
type I from the second order transition for superconductors of
type II. Results of duality arguments \cite{Kleinert82} and Monte
Carlo calculations \cite{Dasgupta81} are in agreement with this
picture, which is also supported by experiment
\cite{Schneider04,Schneider05}.

An important question in calculating critical properties like the critical exponents is their
dependence on the gauge used for the GF. Physically observable quantities should be
independent of the specific gauge used. This has been shown in one loop order \cite{Kang74}.
However non observable quantities might depend on the gauge. Thus the critical exponents
$\nu$ or $\alpha$ of the penetration length $\lambda$ (proportional to the correlation
length $\xi$ of the OP
correlation function at the charged FP) and the specific heat respectively
should be independent of gauge, whereas the critical exponent of the static OP correlation
function turns out to be gauge dependent \cite{Kleinert06}.

The issue of dynamical critical properties was less studied.
Experimental investigations give no consistent picture of the
dynamical critical exponent $z$ (for references see Ref. \cite{Aji01}) obtained from measurements
of the electrical conductivity. The values of the dynamical critical exponents found vary
between 1.5 and 2.3.

Theoretical predictions of $z$ are mainly based on using the results of
known universality classes (model A, C, E or model F, for a review see \cite{FoMo06}).
Model A is the simplest model assuming a relaxation equation for the
OP without coupling to other conserved densities, whereas model C
contains a coupling to a diffusion equation for a conserved density (only relevant if the specific heat
is diverging). Models E and F define the universality class of the critical dynamics of the superfluid transition in
$^4$He. Dynamic equations from the vortex-loop model \cite{Lidmar98,Nogueira05} indicate a relation to that
universality class. So far no systematic derivation of dynamical equations with  mode coupling terms derived from
Poisson bracket relations using for instance methods described in \cite{dzyalo} has been performed.

Monte Carlo simulations \cite{Lidmar98} in the limiting cases of large and small
values of the Ginzburg parameter found values of the dynamical exponent of 2.7 and 1.5
respectively.
Using  the vortex model \cite{Aji01}  for the superconducting transition  the dynamic critical exponent
has been calculated analytically in the two limits mentioned above. The resulting exponents were 5/2 and 3/2
respectively and they were related to the dynamic exponents obtained in simulations. A discussion
\cite{discussion}
arised on the reason of the deviation of the Monte Carlo results from the expected value of model A, which
would lead to a dynamical exponent of roughly 2.
Recently in Ref. \cite{Nogueira05} for {\it extreme} type II superconductors, where the {\it uncharged}
FP describes the static behavior, it was argued using scaling and duality arguments that $z=3/2$ exactly.

A concrete dynamical model defined by equations of motion for the
OP and the GF has been presented in Ref. \cite{Lannert04}. Both
equations are of relaxational character and the essential
parameter is the ratio $w=\Gamma_\psi/\Gamma_A$ of the kinetic
coefficients $\Gamma_\psi$ of the OP and $\Gamma_A$ of the GF. A
dynamic FP, $w^\star$ finite, implying {\it strong scaling} with a
common dynamical critical exponent $z$ for the OP and the GF in
the Feynman gauge (one adds a quadratic term in the divergence of
the GF to the static functional) has been found in one loop order.
However the question of the gauge dependence of the dynamical
exponents was not addressed in Ref. \cite{Lannert04}.

Another important issue not considered so far in the discussion of
the dynamical critical behavior concerns the time scales entering
the problem. One has to discriminate between the time scale for
the OP and its dynamic critical exponent $z_\psi$ and the time
scale of the GF and its dynamic critical exponent $z_A$. They are
defined by the characteristic frequencies of the dynamic
correlation functions for the OP and the GF respectively
\begin{equation}
\omega_\psi\sim k^{z_\psi} g_\psi(k\xi) \qquad  \omega_A\sim k^{z_A} g_A(k\xi) \, .
\end{equation}
$g_\psi$ and $g_A$ are scaling functions which are finite and nonzero at the superfluid transition $T_c$.

The dynamic critical exponents are different \cite{remark1} if  {\it weak scaling} holds otherwise
one says the dynamic correlation functions obey {\it strong scaling}. So far in the discussion the last case was assumed.
Even if for the uncharged FP the dynamics would be in the universality class of model E, it is the {\it weak scaling}
FP which is stable \cite{dedominicispeliti78dohm91}.

Only one  of these dynamical exponents is observable namely $z_A$ entering the frequency
dependent diverging electric conductivity
\begin{equation}
\sigma (\xi ,k=0,\omega)\sim\xi^{(z_A-2+\eta_A)},
\end{equation}
$\eta_A=4-d$ is the anomalous dimension of the GF independent of the gauge \cite{Herbut96}. This implies that the dynamical
critical exponent $z_A$ has to be gauge independent. No such condition holds for $z_\psi$.
If therefore it turns out that $z_\psi$ is gauge dependent {\it strong scaling} cannot hold.

In this paper we apply the field-theoretical RG approach in the minimal subtraction scheme and
dimensional regularization to  the dynamical model of Ref. \cite{Lannert04} in order to investigate
the gauge dependence of the critical dynamics. It turns out that the dynamical exponent $z_\psi$
of the OP is gauge dependent and a gauge independent value for $z_A$ is only obtained in the
{\it weak scaling} FP $w^\star\to \infty$, which however has been found unstable. Therefore
one has to question the physical relevance of the one loop result. This goes together with the problems in
statics of the Ginzburg-Landau model at the one loop order level.

\section{Model}
Static critical properties of charged superconductors in $d$
dimensional space are described by the Abelian Higgs model with the static
functional \cite{Halperin74}:
\begin{eqnarray}\label{hamilt}
{\mathcal H}&=&\int d^d x \Big \{
 \frac{1}{2}{\mathring r} |\vec{\psi}_0|^2
+\frac{1}{2}\sum_{i=1}^{n/2}|(\mbox{\boldmath$\nabla$}-i{\mathring e}\mbox{\boldmath$A$}_0)\psi_{0,i}|^2
+ \frac{\mathring{u}}{4!}(|\vec{\psi}_0|^2)^2+
\frac{1}{2}(\mbox{\boldmath$\nabla$}\times\mbox{\boldmath$A$}_0)^2
 \Big \}.
\end{eqnarray}
with the complex $n$-component OP $\vec{\psi}_0$ (generalized to $n/2$-components, the superconductor
being the case $n=2$) and the $d$-dimensional GF {\boldmath$A$$_0$}. The bare parameter
$\mathring r$ is proportional to the distance from $T_c$ and the bare coupling
${\mathring e}$ is the effective charge. $\mathring u$ is the usual fourth order OP coupling.
In all subsequent calculations we add to the static functional the contribution
\begin{equation}
{\mathcal
H}=-\frac{1}{2\mathring{\varsigma}}(\mbox{\boldmath$\nabla$}\cdot\mbox{\boldmath$A$}_0),
\end{equation}
which allows to choose the gauge.
In the limit $\mathring{\varsigma}\to 0$ the Landau gauge is recovered, while
for system with $\mathring{\varsigma}\to 1$ the Feynman gauge is achieved.

The OP and  the GF are nonconserved quantities thus their dynamical behavior may be described by
relaxation equations. Such a set of equations of motion has been suggested \cite{Lannert04}:
\begin{eqnarray}\label{eq_mov}
\frac{\partial {\psi_{0,i}}}{\partial t}&=&-2\mathring{\Gamma}_{\psi}
\frac{\delta {\mathcal H}}{\delta {\psi^+_{0,i}}}+{\theta}_{i}; \nonumber\\
\frac{\partial {\psi^+_{0,i}}}{\partial t}&=&-2\mathring{\Gamma}_{\psi}
\frac{\delta {\mathcal H}}{\delta {\psi_{0,i}}}+\theta^+_{i}; \\
\frac{\partial {A_{0,\alpha}}}{\partial t}&=&-\mathring{\Gamma}_{A}
\frac{\delta {\mathcal H}}{\delta {A_{0,\alpha}}}+{\theta_{{\alpha}}}. \nonumber
\end{eqnarray}

The OP relaxes with the kinetic coefficient $\Gamma_{\psi}$ while
the GF relaxes with the kinetic coefficient $\Gamma_A$. The
stochastic forces in (\ref{eq_mov}) are related to the kinetic
coefficients and satisfy the relations:
\begin{eqnarray}\label{relat}
<{\theta}_{i}({\bf x},t){\theta}^+_{j}({\bf x}',t')>&=&4\mathring{\Gamma}_{\psi}
\delta({\bf x}-{\bf x}')\delta(t-t')\delta_{ij},
\nonumber\\
<{\theta}_{i}({\bf x},t)>&=&0,
\\
 <{\theta}_{\alpha}({\bf x},t){\theta}_{\beta}({\bf x}',t')>&=&2\mathring{\Gamma}_{A}
 \delta({\bf x}-{\bf x}')\delta(t-t')\delta_{\alpha\beta}, \nonumber \\
<{\theta}_{\alpha}({\bf x},t)>&=&0 \, .  \nonumber
\end{eqnarray}
The indices $i,j$ adopt the values $1,\dots,n/2$, while the indices $\alpha,\beta$ adopt the values $1,\dots,d$.

As it was pointed out in Ref. \cite{Lannert04} the equation for the GF in
(\ref{eq_mov}) can be derived from Maxwell's equations in their low-frequency form. In this case the
inverse transverse coefficient  $\mathring{\Gamma}_A^{-1}$ for the GF can be
identified with the bare normal electrical conductivity.

We  study this critical  dynamics by applying the
Bausch-Janssen-Wagner approach \cite{Bausch76} of dynamical
field-theoretical RG. In this approach, the critical behavior is
considered on the basis of long-distance and long-time properties of
the  Lagrangian incorporating the features defined by the dynamical equations of
the model. In order to keep the powers in the interaction terms low an
auxiliary OP density $\vec{{\tilde{\psi}}}$ and an auxiliary GF ${\tilde{{\boldmath A}}}$
have been introduced. The dynamic equations defined by (\ref{hamilt})-(\ref{relat})
are then described by the unrenormalized Lagrangian
${\mathcal L}={\mathcal L}_0+{\mathcal L}_1$ separated into a Gaussian part




\begin{eqnarray}\label{Lagrangian0}
{\mathcal L}_0&=&\int d^dx dt \Big[
-4\mathring{\Gamma}_{\psi}\sum_{i=1}^{n/2}{\tilde{\psi}^+_{0,i}}{\tilde{\psi}_{0,i}}-
2\mathring{\Gamma}_{A}\tilde{\mbox{\boldmath$A$}}_0^2+\sum_{i=1}^{n/2}\tilde{\psi}^+_{0,i}
\left(\frac{\partial}{\partial
t}+\mathring{\Gamma}_{\psi}({\mathring
r}-\Delta)\right)\psi_{0,i}+\nonumber\\&&
\sum_{i=1}^{n/2}\tilde{\psi}_{0,i}\left(\frac{\partial}{\partial
t}{+}\mathring{\Gamma}_{\psi}({\mathring
r}{-}\Delta)\right)\psi^+_{0,i}
+\tilde{\mbox{\boldmath$A$}}_0\cdot\left(\frac{\partial\mbox{\boldmath$A$}_0}{\partial
t }+\mathring{\Gamma}_A\Big(\mbox{\boldmath$\nabla$}
(1-\frac{1}{\mathring{\varsigma}})(\mbox{\boldmath$\nabla$}\cdot\mbox{\boldmath$A$}_0)
-\Delta\mbox{\boldmath$A$}_0\Big)\right)\Big],
\end{eqnarray}

and an interaction part

\begin{eqnarray}\label{Lagrangian1}
{\mathcal L}_1&=&\int d^dx dt \Big[
\mathring{\Gamma}_{\psi}\frac{\mathring{u}}{3!}|\psi|^2\sum_{i=1}^{n/2}(\tilde{\psi}_{0,i}^+\psi_{0,i}+
\tilde{\psi}_{0,i}\psi^+_{0,i})+2\mathring{\Gamma}_{\psi}i\mathring{e}\mbox{\boldmath$A$}_0\cdot\sum_{i=1}^{n/2}(\tilde{\psi}_{0,i}^+
\mbox{\boldmath$\nabla$}\psi_{0,i}-\tilde{\psi}_{0,i}\mbox{\boldmath$\nabla$}\psi^+_{0,i})
+\nonumber\\&& +\mathring{\Gamma}_{\psi}i\mathring{e}(\mbox{\boldmath$\nabla$}\cdot\mbox{\boldmath$A$}_0)\sum_{i=1}^{n/2}(\tilde{\psi}_{0,i}^+\psi_{0,i}-
\tilde{\psi}_{0,i}\psi^+_{0,i})+\mathring{\Gamma}_A\frac{1}{2}i\mathring{e}\tilde{\mbox{\boldmath$A$}}_0\cdot
\sum_{i=1}^{n/2}(\psi_{0,i}\mbox{\boldmath$\nabla$}\psi^+_{0,i}-\psi^+_{0,i}\mbox{\boldmath$\nabla$}\psi_{0,i})+\nonumber\\&&
\mathring{\Gamma}_{\psi}{\mathring{e}^2}\mbox{\boldmath$A$}_0^2\sum_{i=1}^{n/2}(\tilde{\psi}_{0,i}^+\psi_{0,i}+
\tilde{\psi}_{0,i}\psi^+_{0,i})+\mathring{\Gamma}_A
\mathring{e}^2\tilde{\mbox{\boldmath$A$}}_0\cdot\mbox{\boldmath$A$}_0|\psi|^2\Big].
\end{eqnarray}

\section{Perturbative expansion and renormalization}

\subsection{Vertex functions}
In order to proceed the vertex functions for the OP and the GF are calculated in one loop expansion.
We keep the general structure of the two point vertex functions \cite{FoMo06} and separate
the static contributions. The   calculation leads to the following general form of the OP vertex function
\begin{eqnarray} \label{psistructure}
\mathring{\Gamma}_{\psi\tilde{\psi}^+}&=&-i\omega\mathring{\Omega}_{\psi\tilde{\psi}^+}+2\mathring{\Gamma}_{\psi}
\mathring{\Gamma}^{st}_{\psi{\psi^+}} \, .
\end{eqnarray}
In one loop order the dynamical OP function $\mathring{\Omega}_{\psi\tilde{\psi^+}}$ reads
\begin{eqnarray}
\mathring{\Omega}_{\psi\tilde{\psi}^+}&=&1+4\mathring{e}^2\mathring{\Gamma}_{\psi}\int
\frac{1}{(r{+}(k{+}k')^2)k'^2(-i\omega{+}\mathring{\Gamma}_{\psi}(\mathring{r}{+}(k{+}k')^2)+\mathring{\Gamma}_A
k'^2)}\left(k^2-\frac{(kk')^2}{k'^2}\right)\nonumber\\
&&+\mathring{e}^2\mathring{\Gamma}_{\psi}\mathring{\varsigma}\int
\frac{1}{(\mathring{r}{+}(k{+}k')^2)k'^2(-i\omega\mathring{\varsigma}{+}\mathring{\varsigma}
\mathring{\Gamma}_{\psi}(r{+}(k{+}k')^2){+}\mathring{\Gamma}_A
k'^2)}\left(\frac{((2k{+}k')k')^2}{k'^2}\right),
\end{eqnarray}
whereas  the static OP vertex function is given by
\begin{eqnarray}
\mathring{\Gamma}^{st}_{\psi{\psi^+}}&=&\frac{1}{2}\left\{\mathring{r}+k^2{+}\frac{n+2}{6}\mathring{u}\int\frac{1}{(\mathring{r}
+k'^2)}{+}\mathring{e}^2(d-1+\mathring{\varsigma})\int\frac{1}{k'^2}{-}\right. \nonumber\\
&&\left. 4\mathring{e}^2\int
\frac{1}{(r+(k+k')^2)k'^2}\left(k^2-\frac{(kk')^2}{k'^2}{+}\mathring{\varsigma}\frac{((2k+k')k')^2}{k'^2}\right)\right\}
 \, .
\end{eqnarray}
The same general structure holds for the GF vertex function
\begin{eqnarray} \label{Astructure}
\mathring{\Gamma}^{\alpha\beta}_{A\tilde{A}}&=&-i\omega\mathring{\Omega}^{\alpha\beta}_{A\tilde{A}}+
\mathring{\Gamma}_A{\mathring\Gamma}_{A{A}}^{{st}\,{\alpha\beta}},
\end{eqnarray}
with the one loop expression
\begin{eqnarray}
\mathring{\Omega}^{\alpha\beta}_{A\tilde{A}}&=&
\delta^{\alpha\beta}+\mathring{\Gamma}_{A}\frac{n}{2}\int
\frac{(k+2k')^{\alpha}(k+2k')^{\beta}}{(\mathring{r}+k'^2)(\mathring{r}+(k{+}k')^2)[-i\omega+
\mathring{\Gamma}_{\psi}(\mathring{r}+k'^2)+\mathring{\Gamma}_{\psi}(r+(k{+}k')^2)]},
\end{eqnarray}
for the dynamic GF function and
\begin{eqnarray}
\mathring{\Gamma}^{{st} \,{\alpha\beta}}_{A{A}}&=&
k^2\left(\delta^{\alpha\beta}-\frac{k^{\alpha}k^{\beta}}{k^2}
+\frac{1}{{\mathring{\varsigma}}}\frac{k^{\alpha}k^{\beta}}{k^2}\right)\nonumber\\
&&+\ n\mathring{e}^2\int\frac{1}
{\mathring{r}+k'^2}\left(\delta_{\alpha\beta} -\frac{1}{2}
\frac{(k+2k')^{\alpha}(k+2k')^{\beta}}{\mathring{r}+(k+k')^2}\right),
\end{eqnarray}
for the GF static vertex function.

\subsection{Renormalization and field theoretic functions}

In order to get finite results we perform the renormalization
within the minimal subtraction scheme introducing renormalization
factors leading to the renormalized parameters. The
renormalization factors of the GLW part of the static functional
are introduced as usual
\begin{equation}
\psi_{0,i}=Z^{1/2}_{\psi}\psi_i, \qquad \mathring{r}-\mathring{r}_c=Z_\psi^{-1}Z_rr \, ,
\qquad \mathring{u}={\kappa}^{\varepsilon}{\mathcal A}^{-1}_dZ_uZ^{-2}_{\psi}u  \, ,
\end{equation}
where the shift of the phase transition temperature  $\mathring{r}_c$ has been taken into account.
$\kappa$ represents the free wave vector scale and $\varepsilon=4-d$. We also introduced the usual geometric factor
\[{\mathcal A}_d=\Gamma\left(1-\frac{\epsilon}{2}\right)\Gamma\left(1+\frac{\epsilon}{2}\right)
\frac{\Omega_d}{(2\pi)^d}\,, \]  where $\Omega_d$ is the surface
of the $d$-dimensional unit sphere. The additional renormalization
factors due to the presence of the GF and its coupling to the OP
are introduced as
\begin{equation} \label{renorm}
\qquad A_{0,\alpha}=Z_A^{1/2}A_{\alpha}, \qquad
\mathring{e}^2={\kappa}^{\varepsilon}{\mathcal A}^{-1}_dZ_{e^2}Z_{\psi}^{-1}Z_A^{-1}e^2 \, ,
\qquad \mathring{\varsigma}=Z^{-1}_{\varsigma}Z_{A}\varsigma \, .
\end{equation}
From Ward identities one derives the relations \cite{Kang74}
\begin{equation} \label{Zrel}
Z_{e^2}=Z_{\psi}, \qquad  Z_{\varsigma}=1 \, .
\end{equation}
which show that only one additional (for the GF) renormalization constant with respect to the GLW model appears.

In dynamics two additional renormalization factors are needed which are
\begin{equation}
\tilde{\psi}_{0,i}=Z^{1/2}_{\tilde{\psi}}\tilde{\psi}_i,
\qquad \tilde{A}_{0,\alpha}=Z_{\tilde{A}}^{1/2}\tilde{A}_{\alpha}  \, .
\end{equation}
For the kinetic coefficients no new factors are necessary
\begin{eqnarray}    \label{Zgamma1}
\mathring{\Gamma}_{\psi}&=&Z_{\Gamma_\psi}\Gamma_{\psi}=Z^{1/2}_{\psi}Z^{-1/2}_{\tilde{\psi}}\Gamma_{\psi},
\\  \label{Zgamma2}
\mathring{\Gamma}_{A}&=&Z_{\Gamma_A}\Gamma_A=Z^{1/2}_{A}
Z_{\tilde{A}}^{-1/2}\Gamma_A,
\end{eqnarray}
where the second relations are due to the structure of the vertex functions (see (\ref{psistructure}) and (\ref{Astructure})).

The  renormalization factors calculated in one loop order for the statics are
\begin{eqnarray}    \label{Zstat1}
Z_{\psi}=1+(3-\varsigma)\frac{e^2}{\varepsilon}, \qquad
Z_{A}=1-n\frac{e^2}{6\varepsilon},
\end{eqnarray}
\begin{equation}  \label{Zstat2}
Z_r=1+\frac{n+2}{6}\frac{u}{\varepsilon}-\varsigma\frac{e^2}{\varepsilon},\qquad
Z_{u}=1+\frac{n+8}{6}\frac{u}{\varepsilon}+\frac{18}{u}\frac{e^4}{\varepsilon}-2\varsigma\frac{e^2}{\varepsilon},
\end{equation}
as presented in \cite{Kolnberger90}.  The dynamic renormalization factors are
\begin{eqnarray} \label{Zdyn}
Z_{\tilde\psi}&=&1-(3-\varsigma)\frac{e^2}{\varepsilon}-2\frac{e^2\varsigma}{\varepsilon}\frac{w}{1+w},  \\
Z_{\tilde
A}&=&1+n\frac{e^2}{2\varepsilon}\left(\frac{1}{3}-\frac{1}{w}\right),
\end{eqnarray}
where we have introduced the time scale ratio
\begin{equation} \label{wdef}
w=\frac{\Gamma_{\psi}}{\Gamma_A},
\end{equation}
whose FP value $w^\star$ determines if {\it strong} ($w^\star\ne 0 \ \mbox{or}\ \infty$) or {\it weak scaling}
($w^\star= 0\ \mbox{or}\ \infty$) holds.

From the $Z$-factors one obtains the $\zeta$-functions leading to the $\beta$-functions which determine the FPs.
The critical exponents describing the critical properties are then expressed by the values of the $\zeta$-functions at
the stable FPs. We use the unified definition
\begin{equation}\label{def_z}
\zeta_{a}(\{\alpha\},\varsigma)=-\frac{d\ln Z_{a}}{d \ln \kappa} \, .
\end{equation}
for all  $\zeta$-functions, where $a$ denotes either any model parameter $\{\alpha\}=\{u,e^2,\Gamma_{\psi},\Gamma_A,w\}$ or
any density $\psi,\tilde\psi,A,\tilde A$.
The static $\zeta$-functions following from (\ref{Zstat1}) and (\ref{Zstat2}) read
\begin{eqnarray}
\zeta_{\psi}&=&(3-\varsigma){e^2} \, , \qquad \zeta_{A}=-n\frac{e^2}{6} \, , \\
\zeta_r&=&\frac{n+2}{6}u-\varsigma e^2 \, , \qquad
\zeta_{e^2}=(3-\varsigma){e^2}  \, ,  \\
\zeta_u&=&\frac{n+8}{6}u-2\varsigma e^2+18\frac{e^4}{u}  \, .
\end{eqnarray}
The $\zeta$-function of the gauge parameter follows from (\ref{renorm}) and (\ref{Zrel}) as
$\zeta_{\varsigma}=0$.
The dynamic $\zeta$-functions follow from (\ref{Zdyn}). They are
\begin{equation}  \label{zetatilde}
\zeta_{\tilde\psi}=-(3-\varsigma){e^2}-2e^2\varsigma\frac{w}{1+w}\,,\quad
\zeta_{\tilde
A}=n\frac{e^2}{2}\left(\frac{1}{3}-\frac{1}{w}\right)   \, ,
\end{equation}
and the $\zeta$-functions for the kinetic coefficients are obtained from the above relations  (\ref{Zgamma1})
and (\ref{Zgamma2})
\begin{equation} \label{relaxzeta}
\zeta_{\Gamma_{\psi}}=\frac{1}{2}(\zeta_{\psi}-\zeta_{\tilde\psi}),
\qquad \zeta_{\Gamma_{A}}=\frac{1}{2}(\zeta_{A}-\zeta_{\tilde A}).
\end{equation}
Inserting the one loop results (\ref{zetatilde}) they read
\begin{equation}
\zeta_{\Gamma_{\psi}}=e^2\left(3-\frac{\varsigma}{1+w}\right),
\qquad
\zeta_{\Gamma_{A}}=-n\frac{e^2}{2}\left(\frac{1}{3}-\frac{1}{2w}\right).
\end{equation}
According to  Eqs. (\ref{wdef}) and (\ref{relaxzeta}),  the $\zeta$-function of the time scale ratio $w$
is then found as
\begin{equation}\label{zWW}
\zeta_{w}=\zeta_{\Gamma_{\psi}}-{\zeta_{\Gamma_A}}=e^2\left(3-\frac{\varsigma}{1+w}+
\frac{n}{2}\left(\frac{1}{3}-\frac{1}{2w}\right)\right) \, .
\end{equation}

\section{Fixed points and exponents}

The behavior of the model parameters under renormalization is
described by the flow equations
\begin{equation}\label{fl}
\ell\frac{d\alpha_i}{d \ell}=\beta_{\alpha_i}(\{\alpha\},\varsigma) \,,
\end{equation}
where the right hand sides of the equations are determined by appropriate
$\beta$-functions (the index $i$ runs over the set of parameters).
The $\beta$-functions are generally defined as
\begin{equation}\label{bet}
\beta_{\alpha_i}(\{\alpha_i\},\varsigma)=\alpha_i\left[-c_i-p_i\zeta_\psi-q_i\zeta_A+\zeta_{\alpha_i}\right],
\end{equation}
The coefficients $c_i$, $p_i$ and $q_i$ follow from the general renormalization of $\alpha_i$ as used above
\begin{equation}
\mathring{\alpha}_i={\kappa}^{c_i}{\mathcal A}^{-(p_i+q_i)/2}_dZ_{\psi}^{-p_i}Z_A^{-q_i}Z_{\alpha_i}\alpha_i  \, .
\end{equation}
The FPs $\{\alpha^\star\}$ are defined by the zeros of the right hand sides of Eqs. (\ref{fl}).
A FP is stable if all stability exponents $\omega_i$ are positive. The
stability exponents $\omega_i$ are defined by the eigenvalues of the
matrix $\partial\beta_{\alpha_i}/\partial\alpha_j\vert_{\{\alpha\}=\{\alpha^\star\}}$.

\subsection{Statics}
The one-loop static $\beta$-functions appear to be independent
from the gauge determined by the parameter $\varsigma$ and are the same as
in Ref. \cite{Halperin74}.

\begin{equation}
\beta_{u}=\left(-\varepsilon
u-6e^2u+\frac{n+8}{6}u^2+18e^4\right), \qquad
\beta_{e^2}=e^2\left(-\varepsilon+\frac{n}{6}e^2\right).
\end{equation}
These  $\beta$-functions have four FPs: (i) the Gaussian FP with
$u^\star=e^{\star 2}=0$, (ii) the FP of the uncharged XY-model
($u^\star=6\varepsilon/(n+8),e^\star=0$), (iii) the tricritical FP
with $u^\star=0,e^{\star 2}=6\varepsilon/n$ and (iv) the charged
FP with both couplings  $u^\star\not=0$ and $e^{\star
2}=6\varepsilon/n$ nonzero. In one loop order the charged FP
exists only for $n>365.9$ \cite{Halperin74}. However in higher
loop order and using summation procedures a charged FP is also
found for $n=2$ \cite{FolkHolovatch96}. In one loop order for
$n<365.9$ starting with $e^2\not=0$ and $u\not=0$ the flow escapes
to $e^2\to e^{\star 2}$ and $u\to\infty$.

Critical exponents are calculated at the stable accessible FP.
Expressions for the correlation length critical exponent $\nu$, the specific heat exponent $\alpha$ and the
pair correlation function critical exponent $\eta$ are as follows:
\begin{eqnarray}  \label{nu}
\nu^{-1}&=&2-\zeta_r(u^\star,e^\star,\varsigma)+\zeta_\psi(u^\star,e^\star,\varsigma)=2-\frac{n+2}{6}u^\star+3e^{\star 2}; \\
\alpha&=&2-d\nu ; \\  \label{eta}
\eta&=&\zeta_\psi(u^\star,e^\star,\varsigma)=(3-\varsigma)e^{\star 2}.
\end{eqnarray}
All other static critical exponents may be found from scaling
relations. Since the FP values $u^\star$ and $e^\star$ are gauge
independent this is also valid for $\nu$ and $\alpha$ although the
gauge $\varsigma$ appears in the $\zeta$-functions Eq. (\ref{nu})
explicitly. In contrast this dependence on the gauge remains in
$\eta$ (see Eq. (\ref{eta})). All the other exponents become gauge
dependent since they are in any case related to $\eta$ by the
scaling laws.  However note that only at the charged FP ($e^{\star
2}\ne 0$) a gauge dependence enters the $\zeta$-functions. This
holds in all orders of the loop expansion due to the structure of
the interaction part of the Lagrangian (\ref{Lagrangian1}).

From the renormalization of $\varsigma$ in Equ. (\ref{renorm}) the flow of the gauge parameter reads
\begin{equation}
\ell\frac{d \varsigma}{d \ell}=\varsigma \zeta_A
=\varsigma^2\frac{n}{6}e^2,
\end{equation}
reaching zero at the charged FP. This means only the transverse gauge is invariant under renormalization.

\subsection{Dynamics}

The FP values for the time scale ratio can be found from the zeros of the corresponding
$\beta$-function
\begin{equation}
\beta_{w}=w(\zeta_{\Gamma_{\psi}}-\zeta_{\Gamma_A})=e^2w\left(3-\varsigma
+\varsigma\frac{w}{1+w}+\frac{n}{2}\left(\frac{1}{3}-\frac{1}{2w}\right)\right)
\, .
\end{equation}
These zeros depend in one loop order on the value of the minimal coupling (charge) $e$ at the stable static FP.
Only for non-zero charge a specific value for the possible FP is found in this order ($w$ has to be positive)
\begin{equation}
{w^\star}=\frac{n-12(3-\varsigma)+\sqrt{(n-12(3-\varsigma))^2+24n(18+n)}}{4(18+n)}
\end{equation}
The corresponding stability exponent $\omega_w$
\begin{equation}
\omega_w=\frac{\partial \beta}{\partial w} =e^{\star2}w^\star \left(\frac{\varsigma}{(1+w^\star)^2}+
\frac{n}{4w^{\star2}}\right)
\end{equation}
turns out to be positive and thus the finite FP is stable. Another FP is the
infinite FP. In order to study its stability we introduce
\begin{equation}
\rho=\frac{w}{1+w}
\end{equation}
mapping the range of $w$ into the interval $[0,1]$. The corresponding $\beta$-function reads
\begin{equation}
\beta_\rho=\rho(1-\rho)e^2\left(3-\varsigma+\varsigma\rho+\frac{n}{2}\left(\frac{1}{3}-\frac{1-\rho}{2\rho}\right)\right)=
(1-\rho)e^2\left((3-\varsigma)\rho+\varsigma\rho^2+5\frac{n}{12}\rho-\frac{n}{4}\right).
\end{equation}
The FP $\rho^\star=1$ corresponding to the infinite FP $w^{\star -1}=0$ however
is unstable with the stability exponent
$\omega_{\rho}=-(1+18/n)\varepsilon$.

The dynamical critical exponents for the OP and the GF are calculated from
\begin{equation}
z_\psi=2+\zeta_{\Gamma_\psi}(u^\star,e^\star,w^\star,\varsigma), \qquad
z_A=2+\zeta_{\Gamma_A}(u^\star,e^\star,w^\star) \, .
\end{equation}
Whereas an explicit gauge dependence is found in $\zeta_{\Gamma_\psi}$ this is not the case for
$\zeta_{\Gamma_A}$ at least in one loop order. It is expected that this property holds in all orders.
At the charged FP ($e^{\star 2}=6\varepsilon/n$) in general the exponents take the values
\begin{eqnarray}
z_{\psi}&=&2+\frac{18}{n}\varepsilon-\varsigma\frac{6}{n}\ \frac{\varepsilon}{1+w^\star};\\
z_A&=&2-\varepsilon+\frac{3\varepsilon}{2w^\star}.
\end{eqnarray}
Since $w^\star$ depends on the gauge for the finite FP, {\it the dynamical exponent of the GF, $z_A$,
depends on the gauge in contradiction to the physical requirement that observable quantities should
be gauge independent.}
Only in the case of the infinite FP a gauge independent value is possible. Then
the OP exponent would be finite
\begin{equation}
z_\psi=2+\frac{18}{n}\varepsilon,
\end{equation}
and also independent of the gauge but different from $z_A$. However the infinite FP is not stable
in one loop order. It should be noted that for the gauge $\varsigma=1$ the results of \cite{Lannert04}
are reproduced. It should also be remarked that in one loop order the calculation of the field theoretic
functions is simpler in the transverse gauge ($\varsigma=0$) due to the observation that less
graphical contributions in the loop expansion are nonzero \cite{Kang74}. The gauge independence of the static
FP values for the minimal coupling and fourth order coupling has not been proven to our knowledge.
The finite FP value of the dynamical time scale $w$ is already gauge dependent in one loop order, as has been
shown here, leading to a gauge dependent critical dynamic exponent of the GF. This in our opinion would hold
for every order for the  {\bf finite nonzero} FP value $w^\star$. At the uncharged FP both dynamical
exponents would be independent of the gauge.

\section{Conclusion and Outlook}

We have demonstrated for the dynamical model suggested by Lannert,
Vishveshwara and Fisher \cite{Lannert04} that the OP dynamical
exponent is gauge dependent. Therefore also the exponent for the
divergence of the electric conductivity is gauge dependent at the
stable {\it strong scaling} FP where the dynamical critical
exponents of the OP, $z_\psi$, and the GF, $z_A$, are the same. A
way out in a higher loop order calculation is only possible when
the stability of the {\it strong scaling} FP is changed and the
infinite {\it weak scaling} FP becomes the stable one. As one
knows from other examples one loop order calculations may not be
conclusive and lead to results which have to be taken with care.
In higher order perturbation expansion the stability of the FPs
may be changed. A two loop calculation may clarify the situation.
Recent progress \cite{Canet06} in the nonperturbative version of
dynamic renormalization theory applied to this more complicated
model would be also worthwhile.

\section*{Acknowledgements}%

This work was supported by Fonds zur F\"orderung der
wissenschaftlichen Forschung under Project No. P16574


\begin{thebibliography}{50}
\bibitem{remark1}  In statics a difference in the scaling of lengths appears depending on
which FP is stable. At the XY FP (FP value of the coupling to the
GF is zero) the OP correlation length $\xi$ scales with the
exponent $\nu$, whereas the penetration length $\lambda$ with
$\nu/2$. At the charged FP (FP value of the coupling to the GF is
non-zero) both lengths scale with the exponent $\nu$ (see
\cite{Herbut96}.
\end{thebibliography}
\end{document}